\documentclass[aip,jmp,reprint,apl]{revtex4-1}
\usepackage{color}
\usepackage{graphicx}
\usepackage{dcolumn}
\usepackage{bm}
\usepackage{amsmath}

\begin{document}

\title{2$\mu$m Solid-State Laser Mode-locked By Single-Layer Graphene}
\author{A. A. Lagatsky$^1$,Z. Sun$^2$,T. S. Kulmala$^2$,R. S. Sundaram$^2$,S. Milana$^2$, F. Torrisi$^2$, O. L. Antipov$^3$ \\Y. Lee$^4$, J. H. Ahn$^4$, C. T. A. Brown$^1$, W. Sibbett}
\affiliation{$^1$School of Physics and Astronomy, University of St Andrews, St Andrews, KY16 9SS UK\\
$^2$Department of Engineering,University of Cambridge,Cambridge CB3 0FA,UK\\
$^3$Institute of Applied Physics, Russian Academy of Sciences, Nizhny Novgorod Russia\\
$^4$School of Advanced Materials Science and Engineering and Advanced Institute of Nanotechnology, Sungkyunkwan University, Suwon 440-746, Korea\\}
\author{A.C. Ferrari$^2$}
\email{acf26@eng.cam.ac.uk}
\begin{abstract}
We report a 2$\mu$m ultrafast solid-state Tm:Lu$_2$O$_3$ laser, mode-locked by single-layer graphene, generating transform-limited$\sim$410fs pulses, with a spectral width$\sim$11.1nm at 2067nm. The maximum average output power is 270mW, at a pulse repetition frequency of 110MHz. This is a convenient high-power transform-limited laser at 2$\mu$m for various applications, such as laser surgery and material processing.
\end{abstract}

\maketitle

\noindent Ultrafast lasers operating at$\sim$2$\mu$m are of great interest due to their potential in various applications, e.g. telecoms\cite{Chan_jstqe_2010}, medicine\cite{Bouma_jbo_1998,medical_appl}, material processing\cite{Gattass_np_2008,medical_appl} and environment monitoring\cite{Ebrahim_book}. They can be used for light detection and ranging measurements\cite{Ebrahim_book} and free-space optical communications\cite{Ebrahim_book}, due to the 2-2.5$\mu$m atmospheric transparency window\cite{Ebrahim_book}. Because water (main constituent of human tissue) absorbs more strongly at$\sim$2$\mu$m ($\sim$100/cm)\cite{medical_appl} than at other conventional laser wavelengths (e.g.$\sim$10/cm at$\sim$1.5$\mu$m, and$\sim$1/cm at$\sim$1$\mu$m)\cite{medical_appl}, sources working at$\sim$2$\mu$m are promising for medical diagnostic\cite{medical_appl} and laser surgery\cite{medical_appl}. Currently, the dominant technique for ultrafast pulse generation at 2$\mu$m relies on semiconductor saturable absorber mirrors (SESAMs)\cite{Keller_nature_03,Sibbett_ol_2012}. InGaAsSb quantum-well-based SESAMs have been used to mode-lock Tm,Ho:NaY(WO$_{4}$)$_{2}$\cite{Lagatsky_ol_2010} and Tm:Sc$_{2}$O$_{3}$\cite{Lagatsky_ol_2012} lasers, generating 258fs pulses with 155mW output power at 2$\mu$m\cite{Lagatsky_ol_2010}, and 246fs pulses with 325mW output at 2.1$\mu$m\cite{Lagatsky_ol_2012}. However, SESAMs require complex growth techniques (e.g. molecular beam epitaxy\cite{Keller_nature_03}), often combined with ion implantation\cite{Lagatsky_ol_2010,Lagatsky_ol_2012} to reduce recovery time\cite{Keller_nature_03,Sibbett_ol_2012}.

Nanotubes and graphene have emerged as promising saturable absorbers (SA), due to their low saturation intensity\cite{Hasan_am_2009,Sun_an_10,Sun_pe_12,Bonaccorso_np_10,Popa_apl_2012}, low-cost\cite{Hasan_am_2009} and easy fabrication\cite{Bonaccorso_np_10,Sun_an_10,Torrisi_an_2012}. With nanotubes, broadband operation can be achieved by using a distribution of tube diameters\cite{Wang_nn_2008,Hasan_am_2009}. With graphene, this is intrinsic, due to the gapless linear dispersion of Dirac electrons\cite{Sun_an_10,Bonaccorso_np_10}. Ultrafast pulse generation at 0.8\cite{Baek_ape_2012}, 1\cite{Tan_apl_10}, 1.3\cite{Cho_ol_08} and 1.5$\mu$m\cite{Bonaccorso_np_10,Hasan_am_2009,Sun_pe_12,Sun_an_10,Sun_tun,Bao_afm_09,Hasan_pssb_10,Popa_2010} was demonstrated with graphene-based SAs (GSAs). Ref.\onlinecite{Zhang_oe_2012} reported a 1.94$\mu$m Tm-doped fiber laser mode-locked by a polymer composite with graphene produced by liquid phase exfoliation of graphite \cite{Hernandez_nn_08,Sun_an_10}. Compared to solid-state lasers, fiber lasers have some advantages, such as compact geometry and alignment-free operation. However, their output power is typically very low ($\sim$mW\cite{Nelson1997}) and their output spectrum generally has side-bands\cite{Nelson1997}. Solid-state lasers have the advantage, compared to fibre lasers, of sustaining ultrafast pulses with higher output power (typically$\geq$100mW)\cite{Keller_nature_03,Sibbett_ol_2012} and better pulse quality (e.g. transform-limited with sideband-free profile in the spectral domain\cite{Keller_nature_03,Sibbett_ol_2012}). Therefore, solid-state lasers are of interest for applications requiring high power and good pulse quality, such as industrial material processing\cite{Keller_nature_03} and laser surgery\cite{medical_appl}. Ref.\onlinecite{Liu_lpl_11} used graphene-oxide to mode-lock a 2$\mu$m solid-state Tm:YAlO$_{3}$ laser. However, the output pulse duration was long,$\sim$10ps, due to the lack of intracavity dispersion compensation\cite{Liu_lpl_11}. Also, graphene oxide\cite{stankovich_nat_06,Mattevi_afm_09} is fundamentally different from graphene: it is insulating, with a mixture of sp$^2$/sp$^3$ regions\cite{stankovich_nat_06,Mattevi_afm_09}, with many defects and gap states\cite{Mattevi_afm_09}. Thus it may not offer the same wideband tunability as graphene. A mixture of 1 or 2 graphene layers grown by Chemical Vapor Deposition (CVD) was used to mode-lock a Tm-doped calcium lithium niobium gallium garnet (Tm:CLNGG) laser at 2$\mu$m in Ref.\onlinecite{Ma_ol_2012}. However, compared to 2$\mu$m solid-state lasers mode-locked by SESAMs\cite{Lagatsky_ol_2010,Lagatsky_ol_2012}, the output power was low($\sim$60mW), limited by damage to the mode-locker.

Here we report a single-layer graphene (SLG) mode-locked solid-state Tm:Lu$_2$O$_3$ laser at$\sim$2067nm,with a 270mW average output power. Transform-limited$\sim$410fs pulses are generated using a dispersion-compensated cavity. This is a convenient high-power transform-limited laser at 2$\mu$m for various applications.

\begin{figure}[htb]
\centerline{\includegraphics[width=80mm]{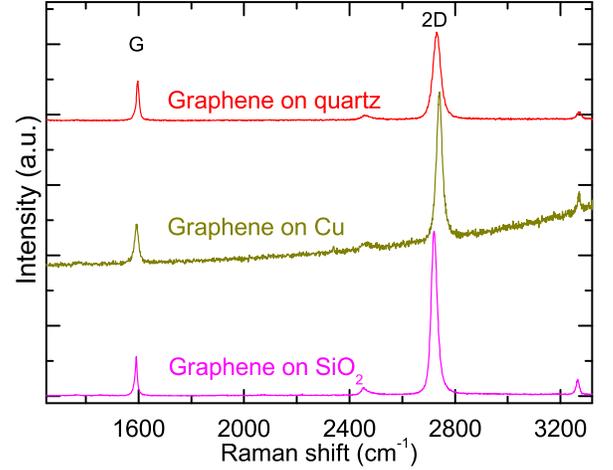}}
\caption{Raman spectra at 457nm for graphene on Cu (before transfer) and after transfer on quartz and SiO$_{2}$/Si}
\label{Raman}
\end{figure}

Our GSA is prepared as follows. SLG is grown by CVD\cite{Li_s_2009,Bae_nn_10}. A$\sim$35$\mu$m thick Cu foil is heated to 1000 $^{o}$C in a quartz tube, with 10sccm H$_{2}$ flow at$\sim$5$\times$10$^{-2}$ torr. The H$_{2}$ flow is maintained for 30mins. This not only reduces the oxidized foil surface, but also extends the graphene grain size. The precursor gas, a H$_{2}$:CH$_4$ mixture with flow ratio 10:15, is injected at a pressure of 4.5$\times$10$^{-1}$ torr for 30mins. The carbon atoms are then adsorbed onto the Cu surface and nucleate SLG via grain propagation\cite{Li_s_2009,Bae_nn_10}. The quality and number of layers are investigated by Raman spectroscopy\cite{Ferrari_prl_06,Cancado_nl_2011}, Fig.\ref{Raman}. At the more common 514nm excitation, the Raman spectrum of CVD graphene on Cu does not show a flat background, due to Cu photoluminescence\cite{Mooradian_prl_1969}. This can be suppressed at 457nm, Fig.1. The spectrum does not show a D peak, indicating the absence of structural defects\cite{Ferrari_prl_06,Cancado_nl_2011,Ferrari_prb_00}. The 2D peak is a single sharp Lorentzian, signature of SLG\cite{Ferrari_prl_06}.

\begin{figure}[htb]
\centerline{\includegraphics[width=80mm]{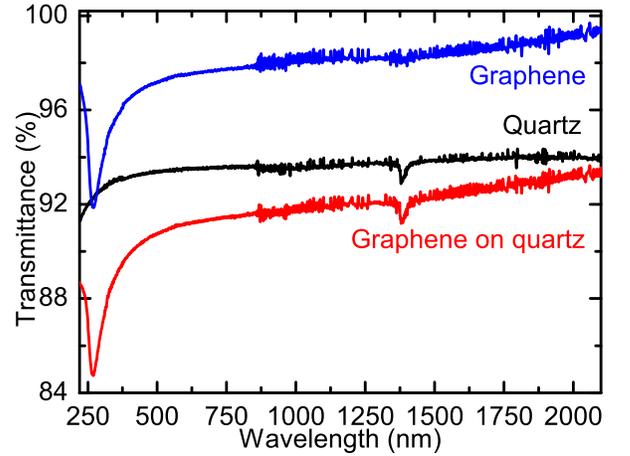}}
\caption{Transmittance of quartz and graphene on quartz. For graphene, this is derived from the transmittance of transferred graphene on quartz divided by that of quartz.}
\label{Absorption}
\end{figure}

We then transfer a 10$\times$10mm$^2$ SLG region onto a quartz substrate (3mm thick) as follows. Poly(methyl methacrylate) (PMMA) is spin-coated on the sample. Cu is then dissolved in a 3\% H$_{2}$O$_{2}$:35\% HCl (3:1 ratio) mixture, further diluted in equal volume of deionized water. The PMMA/graphene/Cu foil is then left floating until all Cu is dissolved. The remaining PMMA/graphene film is cleaned by moving it to a deionized H$_{2}$O bath, a 0.5M HCl bath, and again to a deionized H$_{2}$O bath. Finally, the layer is picked up using the target quartz substrate and left to dry under ambient conditions. After drying, the sample is heated to 180$^{o}$C for 20mins to flatten out any wrinkles\cite{Pirkle_apl_2011}. The PMMA is then dissolved in acetone, leaving SLG on quartz. This is then inspected by optical microscopy, Raman spectroscopy and absorption microscopy. A representative Raman spectrum of the transferred sample is in Fig.\ref{Raman}. After transfer, the 2D peak is still a single sharp Lorentzian, validating that SLG has indeed been transferred. The absence of a D peak proves that no structural defects are induced during this process \cite{Ferrari_prl_06,Cancado_nl_2011,Ferrari_prb_00}. In order to estimate the doping, an analysis of more than 15 measurements with 514nm excitation is carried out. This wavelength is used since most previous literature and correlations were derived at 514nm\cite{Das_nn_2008}. We find that the G peak position, Pos(G), up-shifts$\sim$4cm$^{-1}$ in average after transfer on quartz, whereas the full width at half maximum of the G peak, FWHM(G), decreases from$\sim$17 to$\sim$10.5cm$^{-1}$. Also, the 2D to G intensity and area ratios, I(2D)/I(G); A(2D)/A(G), decrease from 3.2 to 1.6 and 5.8 to 5.3, respectively. This implies an increased p-doping compared to graphene on Cu before transfer\cite{Das_nn_2008,Casiraghi_apl_2007,Pisana_nm_2007}. We estimate the doping for the sample on quartz to be$\sim$10$^{13}$cm$^{-2}$, corresponding to a Fermi level shift$\sim$300/400meV. For comparison, we also transferred on SiO$_{2}$/Si. In this case, the average Pos(G) and FWHM(G) are 1584cm$^{-1}$  and 14cm$^{-1}$, respectively. The average Pos(2D) is 2685cm$^{-1}$, and I(2D)/I(G); A(2D)/A(G) are 3.2 and 7.1, respectively. This indicates a much lower doping, below 100meV. Therefore, we conclude that the doping of our graphene transferred on quartz does not arise from the transfer process itself, but it is most likely due to charge transfer from adsorbates on the substrate\cite{Wang_jpcc_2008,Kong_jpcc_2010}. The transmittance of the transferred SLG on quartz is then measured (Fig.\ref{Absorption}). The band at$\sim$270nm is a signature of the van Hove singularity in the graphene density of states\cite{krav}, while those at$\sim$1.4, 2.2$\mu$m are due to quartz\cite{Agrawal_book_app}. The transmittance in the visible range (e.g. at$\sim$700nm) is$\sim$97.7\% (i.e.,$\sim$2.3\% absorbance), further confirming that the sample is indeed SLG\cite{Nair_s_08}. The absorbance decreases to$\sim$1\% at 2067nm, much lower than the 2.3\% expected for intrinsic SLG. We assign this to doping\cite{Mak_prl_08}. The graphene optical conductivity $\sigma$ at a wavelength $\lambda$ is: $\sigma (\lambda,E_{F},T)=\frac{\pi e^2}{4 h} [ tanh(\frac{ \frac{h c}{\lambda} + 2 E_{F}}{4 k_B T})+ tanh(\frac{\frac{h c}{\lambda} -2 E_{F}}{4 k_B T})] $, as for Ref.\onlinecite{Mak_prl_08}, where T is the temperature, $E_{F}$ the Fermi energy. The transmittance (Tr) is linked to $\sigma$ as:\cite{Mak_prl_08} $Tr\approx 1-\frac{4\pi\sigma}{c} $. By fitting to the measured Tr we derive $E_{F}\sim$350meV, consistent with the Raman estimates.

\begin{figure}[htb]
\centerline{\includegraphics[width=70mm]{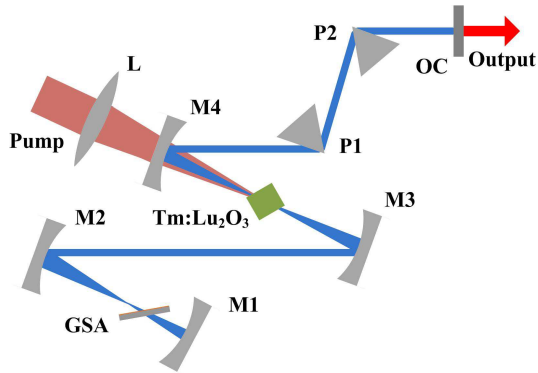}}
\caption{Laser setup. L: lens; M1 with 75mm curvature; M2-M4 with 100mm curvature radii; P1,P2: fused silica prisms.}
\label{Setup}
\end{figure}

The laser setup is shown in Fig.\ref{Setup}. The cavity consists of four plano-concave high-reflectivity (R$>$99.2\% at 2$\mu$m) mirrors (M1-M4) and an output coupler (OC) with 1\% transmittance at 2$\mu$m, and is designed to ensure the best mode-matching between the pump and intra-cavity laser beams. Tm:Lu$_2$O$_3$ ceramic is selected as the gain material because of its high conductivity\cite{Koopmann_apb_2011}, broad emission spectrum ($>$1.9-2.1$\mu$m\cite{Koopmann_apb_2011,Oleg_qe_2011}), high absorption \cite{Koopmann_apb_2011,Oleg_qe_2011} and emission cross-sections \cite{Koopmann_apb_2011,Oleg_qe_2011}, making it suitable for high-power ultrafast pulse generation\cite{Koopmann_apb_2011,Oleg_qe_2011,Lagatsky_oe_2012}. A 5mm long Tm:Lu$_2$O$_3$ ceramic is pumped by a home-made continuous-wave Ti:sapphire laser at 796nm with 2.6W maximum power. A \textit{p}-polarized pump beam is focused into the gain medium via an 80mm focal length lens and a folding mirror (with$>$99\% transmittance at 976nm) to a spot radius of 26$\mu$m (1/e$^2$ intensity), measured in air at the location of the input facet of the ceramic. The GSA is inserted in the cavity between mirrors M1 and M2 at the Brewster's angle, to reduce Fresnel reflection loss (Fig.\ref{Setup}). The laser beam waist radii inside the gain medium and on the GSA are calculated as 32$\times$61$\mu$m$^2$ and 110$\times$158$\mu$m$^2$, respectively, by using the ray matrices method of Ref.\onlinecite{Koechner_book}. A pair of infrared-grade fused silica prisms with 12cm tip-to-tip separation is used to control the intracavity net group velocity dispersion (GVD). Each prism is placed at a minimum deviation to reduce insertion losses. The total round-trip cavity GVD at 2$\mu$m is$\sim$ -2980fs$^2$, due to the insertion of the prisms (glass material dispersion, -113 fs$^2$/mm), the gain medium itself (-15fs$^2$/mm) and the angular dispersion of the prism pair (-1436fs$^2$). The whole cavity length is$\sim$1.35m.
\begin{figure}[htb]
\centerline{\includegraphics[width=80mm]{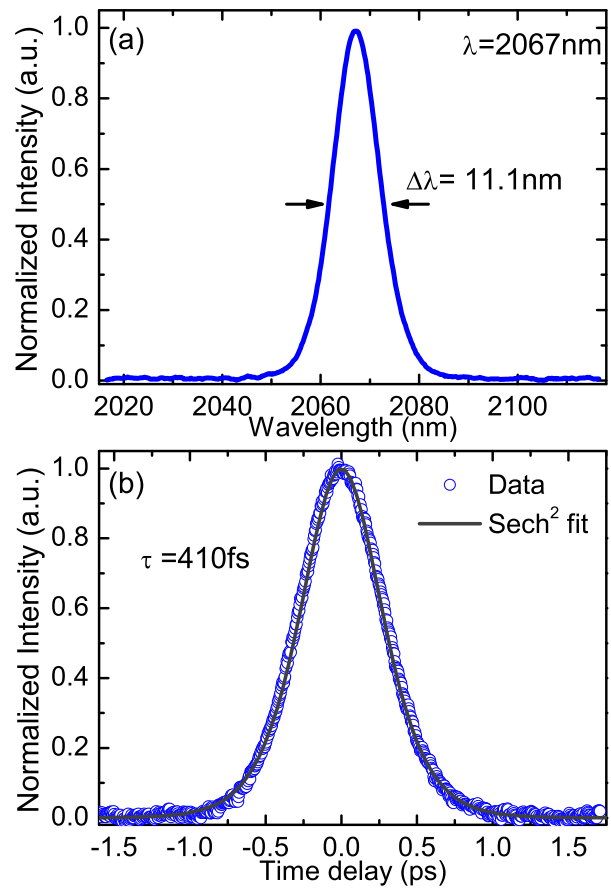}}
\caption{(a) output spectrum,(b)autocorrelation trace.}
\label{OP}
\end{figure}
\begin{figure}[htb]
\centerline{\includegraphics[width=80mm]{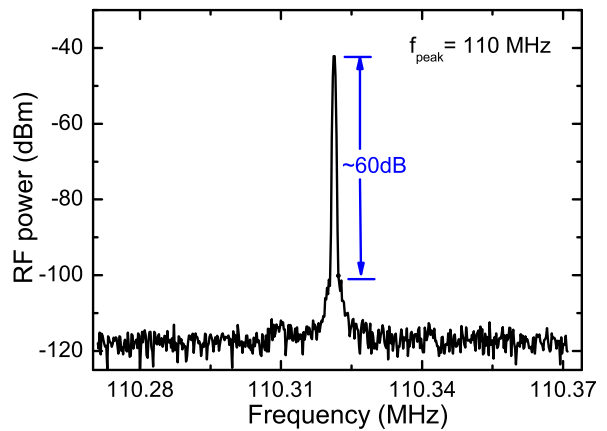}}
\caption{RF spectrum. The resolution bandwidth is 300Hz.}
\label{RF}
\end{figure}

During continuous wave operation (without GSA) the laser produces up to 640mW output power from 1.8W of absorbed pump power at$\sim$2070nm, the lasing threshold being 89mW. After inserting the GSA, the lasing threshold increases to 314mW. Self-starting mode-locking is achieved at 160mW average output power (with$\sim$1.16W absorbed pump power). The maximum average output power is 270mW, while the absorbed pump power is 1.8W. The obtained output power is comparable to that of previous 2$\mu$m SESAMs mode-locked ultrafast solid-state lasers (e.g. 155mW from Tm,Ho:NaY(WO$_{4}$)$_{2}$\cite{Lagatsky_ol_2010}, 325mW from Tm:Sc$_{2}$O$_{3}$\cite{Lagatsky_ol_2012} lasers), but larger than thus far reported for 2$\mu$m nanotube mode-locked Tm-doped solid-state lasers (e.g. 50mW from a Tm:Lu$_{2}$O$_{3}$ laser\cite{Schmidt_oe_2012}) and graphene mode-locked solid-state lasers (e.g. 60mW from a Tm:CLNGG laser\cite{Ma_ol_2012}) in sub-ps regime. The repetition rate is$\sim$110MHz. The corresponding pulse energy is$\sim$2.45nJ, higher than thus far achieved for 2$\mu$m nanotube (e.g.$\sim$0.5nJ\cite{Kieu_ptl_09})\cite{Kivisto_oe_09,Solodyankin_ol_08} and graphene (e.g.$\sim$0.4nJ\cite{Zhang_oe_2012}) mode-locked fiber lasers\cite{Zhang_oe_2012}. Higher output power/energy is possible by increasing pump power, as the output power is limited by the maximum available pump power.

The peak wavelength is 2067nm (Fig.\ref{OP}a). The FWHM bandwidth is$\sim$11.1nm at the maximum average output power. The spectrum has no soliton sidebands, unlike what typical for 2$\mu$m ultrafast fiber lasers\cite{Solodyankin_ol_08,Kieu_ptl_09,Kivisto_oe_09} due to intracavity periodical perturbations\cite{Dennis_jqe_1994}. Fig.\ref{OP}b plots the autocorrelation trace of the output pulses at the maximum average output power. The data are well fitted by a sech$^{2}$ temporal profile, giving a pulse duration$\sim$410fs. This is longer than previously reported from SESAMs and nanotube mode-locked 2$\mu$m solid-state lasers (e.g.$\sim$200fs\cite{Lagatsky_ol_2010,Lagatsky_ol_2012,Schmidt_oe_2012}), but shorter than previous graphene mode-locked 2$\mu$m solid-state lasers (e.g.$\sim$10ps\cite{Liu_lpl_11},$\sim$729fs\cite{Ma_ol_2012}). The pulse duration is much shorter than 2$\mu$m nanotube  (e.g.$\sim$0.75ps\cite{Kieu_ptl_09},$\sim$1.3ps\cite{Solodyankin_ol_08}) and graphene (e.g.$\sim$3.6ps\cite{Zhang_oe_2012}) mode-locked fiber lasers. The time-bandwidth product is 0.319, close to 0.315 expected for transform-limited sech$^{2}$ pulses.

The mode-locking operation stability is studied measuring the radio frequency (RF) spectrum using a fast InGaAs photo-detector ($>$7GHz cut-off) connected to a spectrum analyzer. Fig.\ref{RF} plots the RF spectrum around the fundamental repetition frequency of 110MHz. A signal-to-noise ratio of 60dB (a contrast of 10$^6$) is measured, implying no Q-switching instabilities\cite{Honninger_josa_99}.

In conclusion, we demonstrated a graphene mode-locked solid-state Tm:Lu$_{2}$O$_{3}$ laser at 2$\mu$m, having transform-limited 410fs pulses with$\sim$270mW average output power, and$\sim$110MHz repetition rate. This showcases the potential of graphene for high-power ultrafast solid-state lasers.

We acknowledge funding from the ERC grants NANOPOTS, EU grants RODIN, MEM4WIN, GENIUS, EPSRC grants EP/GO30480/1 and EP/G042357/1, King's college Cambridge, the Royal Academy of Engineering, a Royal Society Wolfson Research Merit Award, and the Cambridge Nokia Research Centre.

\end{document}